\documentclass{article}

\usepackage {graphicx,latexsym}
\usepackage{amsmath,amsthm}
\usepackage{amsfonts}
\usepackage{amssymb}
\usepackage[utf8]{inputenc}
\usepackage{authblk}
\usepackage{caption}
\usepackage{subcaption}
\usepackage{cite}
\usepackage[mathscr]{euscript}

\newcommand{\R}{{\mathbb R}}

\begin{document}

\title{
Towards relational foundations for spacetime quantum physics
}
\author[1]{
Pietro Dall'Olio
\footnote{e-mail: \ttfamily pietro.dallolio@lyon.edu}
}
\author[2]{
José A. Zapata
\footnote{e-mail: \ttfamily zapata@matmor.unam.mx}
}
\affil[1]{
Division of Mathematics and Science, \\
Lyon College, Batesville, Arkansas, USA.
}
\affil[2]{
Centro de Ciencias Matemáticas \\
Universidad Nacional Autónoma de México, 
Morelia, México.
}

\date{}
\maketitle

\begin{abstract} 
Rovelli’s relational interpretation of quantum mechanics tells us that the description of a system in the formalism of quantum mechanics is not an absolute, but it is relative to the observer itself. The interpretation goes further and proposes a set of axioms. 
In standard non relational language, 
one of them states that an observer can only retrieve finite amount information from a system by means of measurement. Our contribution starts with the observation that quantum mechanics, i.e. quantum field theory (QFT) in dimension 1, radically differs from QFT in higher dimensions. In higher dimensions boundary data (or initial data) cannot be specified by means of finitely many measurements. This calls for a notion of measuring scale, which we provide. At a given measuring scale the observer has partial information about the system. Our notion of measuring scale generalizes the one implicitly used in Wilsonian QFT, where at each measuring scale there are effective theories, which may be corrected, and if the theory turns out to be renormalizable the mentioned corrections converge to determine a completely corrected (or renormalized) theory at the given measuring scale.
The notion of a measuring scale is the cornerstone of Wilsonian QFT. This notion tells us that we are not describing a system from an absolute perspective. An effective theory at that scale describes the system with respect to the observer, which may retrieve information from the system by means of measurement in a specific way determined by our notion of measuring scale.
We claim that a relational interpretation of quantum physics for spacetimes of dimension greater than 1 is Wilsonian.

\end{abstract}

\section{Foundations for spacetime quantum physics}

A spacetime context is universally accepted as the most convenient way to set up the foundations of classical physics. 
As far as foundations are concerned, the setting in which time is the only independent variable is considered to be just an auxiliary setting that may be technically useful. There seems to be no reason to go back to a ``time only'' context to address the foundations of quantum physics. 
We observe, however, that in the literature about the foundations of quantum physics there is a remarkable elusion of aspects associated to the spatiotemporal nature of our world. The hypothesis behind this paper is that the foundations of quantum physics should not be confined to having only time as an independent variable. 

Our particular goal is to contribute to export ideas of Relational Quantum Mechanics (RQM) \cite{Rovelli:1996, Rovelli:St2025}, a relational interpretation of quantum mechanics, and complement them as appropriate to give a foundational interpretation which applies to spacetime systems (i.e. fields). Is it a trivial task? Is the essence of quantum physics captured in 1-dimensional spacetimes, or are there key features of quantum physics that can only be appreciated in higher dimensional spacetimes? The contribution of this paper is a ``finite resolution postulate'' for RQP, a Relational Interpretation of Quantum Physics (quantum field theory) constructed as a higher dimensional counterpart of RQM, which brings in the Wilsonian interpretation of QFT \cite{Wilson:I1971, Wilson:II1971} to RQP. 

Wilsonian QFT relies in the concept of scale: At a given scale (energy scale or length scale), a model with a cutoff (using a finite dimensional space of states) can describe the system up to some error. For some systems, the error may be reduced to zero by introducing more degrees of freedom and complicating the model, and the resulting ``completely renormalized model'' correctly describes the system at the given scale. 
Wilsonian renormalization is equivalent to the ``systematic killing of infinities'' \cite{Stueckelberg:1953, Gell-Mann:1954, Weinberg:1960}, but conceptually it is much more enlightening. See \cite{Callan:1970, Wilson:1974, Polchinski:1983} for an elucidative explanation on how the apparently \emph{ad hoc} prescription for curing divergences in particle physics is justified from first principles within the Wilsonian reformulation of the renormalization group.

Renormalization is not included in quantum mechanics textbooks because, in most situations, one can make sense of quantum mechanics without appealing to a renormalization process. On the other hand, QFT without renormalization only describes free theories or exactly solvable models. The reason for this sharp difference is topological: The system may be characterized by initial data given over surfaces of codimension 1. In dimension 1, where time is the only independent variable, the ``initial data surface'' is a single point $t_0 \in \R_t$. Then the space of states of the system is finite dimensional. 
On the other hand in dimension 2 and higher the initial data surface has infinite cardinality, and the resulting space of states of the system is infinite dimensional. 
Thus, renormalization is relevant in dimension 2 and higher, and it is not relevant in dimension 1.

We use the notion of measuring scale proposed in \cite{Manrique:2006} which is appropriate in a relational context. In a laboratory situation, the experimentalists set up an array of measuring devices, filters and control elements which determine the algebra of observables accesible to the experimentalist. That algebra can be seen as a subalgebra of the algebra of observables that could in principle be measured. That subalgebra is the measuring scale \cite{Manrique:2006}. Measuring scales are a partially ordered set, and coarse graining operations are injective algebra morphisms associated to inclusion of observables from coarser to finer scales. A continuum limit is also defined in this implementation of Wilsonian QFT. 

The finite resolution postulate has origin in Wilsonian QFT and also in RQM. We first noticed that the cornerstone of the Wilsonian view of QFT looked compatible with one of the central elements of RQM. Then we brought in the appropriate notion of scale that smoothly brings them together. 

Section \ref{RQM} gives a brief description of Rovelli's Relational Quantum Mechanics. 
Section \ref{RQPisWilsonian} states the problem of finding a relational interpretation of Quantum Field Theory, and provides initial ingredients. In particular the Finite Resolution Postulate is introduced. 
In Section \ref{Measurement/InteractionInRQP} the problem of measurement/interaction is examined from a relational perspective following the formalism of Fewster and Verch \cite{Fewster:2020}. 
A summary and outlook section finishes the article.

\section{A brief description of RQM}
\label{RQM}

The Relational interpretation of Quantum Mechanics (RQM) proposed by Rovelli in 1996 \cite{Rovelli:1996} refines the Copenhagen interpretation of QM as presented in textbooks to (i) emphasize the fact that the key concept of observable of the Copenhagen interpretation corresponds to an action on the system by another system (which is also a quantum system), and (ii) to consider the possibility of having a system described with respect to different reference systems. Paraphrasing Rovelli, RQM allows for different perspectives, whereas the standard Copenhagen interpretation implicitly considers only one. 

RQM also democratizes the status of “the observer”. This does not mean that an atom knows quantum physics and can describe another atom in a certain language. RQM abandons the word “observable” with the purpose of not restricting the agents that can act on the system to agents with consciousness; the term replacing it is “relational variable”. It is a function that describes a property of system ${\mathcal S}$ with respect to system ${\mathcal S}'$, and it also describes a mutual action of systems ${\mathcal S}'$ and ${\mathcal S}$ on each other. 

Two interacting systems ${\mathcal S}$ and ${\mathcal S}'$ do not act on each other continuously; instead they do it in a discrete set of quantum events, and when the action takes place it is, in a certain sense, brusque. 
Relational variables play a dual role. On the one hand, their possible values describe system ${\mathcal S}$ with respect to system ${\mathcal S}'$. On the other hand, they describe the interaction between ${\mathcal S}$ and ${\mathcal S}'$. As usual in Quantum Mechanics, relative variables do not assume definite values, instead, all the possible evaluations coexist until a quantum event associated with that relative variable happens. In that moment, the relative variable takes one of its possible values. 
This central difference between Classical and Quantum Mechanics is emphasized by RQM. In Classical Mechanics, one pretends that the measurement of a system can be done with infinite care, in such a way as not to disturb the system. Quantum Mechanics drops that assumption and assumes the consequences. The result is a formalism that more accurately describes “light systems” (those that are more sensitive to a brusque action). This phenomenon is described by Gamow in the dreams of Mr. Tompkins in a world with a very large $\hbar$. In that world you cannot pat your pet cat: you either miss or you break its neck. This image is appropriate for a description of RQM, it emphasizes that it is not about the nature of measurement, but about the nature of interaction between two systems. 

In \cite{Rovelli:St2025} the properties described in the previous paragraph furnish what is called a “sparse event ontology”. 

According to RQM, QM is about describing interactions between systems in the events when they act on each other. More precisely, given two interacting systems ${\mathcal S}$, $S’$, the goal of QM is to predict the probability distribution that values of relative $S\text-S’$ variables will take at future interaction events given the history of previous values of relative $S\text-S’$ variables. 

An important element of RQM is that the above paragraph talks only about $S\text-S’$ relative variables and not about variables describing S with respect to any other system. For a discussion of Wigner’s friend scenario according to RQM see \cite{Rovelli:1996}. For a discussion regarding the coexistence of different perspectives in the description of a system see \cite{Rovelli:St2025, Adlam:2022} and references therein. For an implementation of a perspective-neutral framework see \cite{delaHamette:2021oex}.

The history of evaluations of relative $S\text-S’$ variables in previous events may be stored in a {\em state} describing ${\mathcal S}$ with respect to $S’$. It is important to notice that, according to RQM, the state is a useful bookkeeping device, but it is not an ontological element of Quantum Mechanics. From this perspective, it is not necessary to include a ``collapse of the wave function'' in the foundations of QM: the element of abrupt change is associated to the ``sparse event ontology''. When a system ${\mathcal S}'$ acts on system ${\mathcal S}$ leading to a quantum event, say at time $t$, in which a relative ${\mathcal S}\text-{\mathcal S}'$ variable, say $V$, takes a given value, say $\lambda$, the bookkeeping device stores that information, and in future events the probabilities for the evaluation of relative variables are {\em conditional} to the fact that at time $t$ variable $V$ evaluated to $\lambda$. 
This point of view does not solve all philosophical problems associated to measurement in QM, but it brings a new angle into the discussion; for more details see 
\cite{Rovelli:St2025} and references therein. 

The ``sparse event ontology'' of RQM brings in the physical question {\em When do quantum events take place?} In \cite{Rovelli:1998} the problem is solved: Quantum events are sharply located in time --they happen at specific times--. QM can be used to calculate the probability density in the real line for an event to happen at each given time. In Section \ref{Measurement/InteractionInRQP} we argue that in higher dimensional spacetimes the picture changes making events fuzzy instead of sharply located.






\section{Relational quantum physics of spacetime systems is Wilsonian}
\label{RQPisWilsonian}

Relational Quantum Physics (RQP) is a relational interpretation of Quantum Field Theory (QFT). It relies on a Lorentzian spacetime background $M$, Minkowski spacetime is a good example to have in mind. The independent variables live in $M$, and the notions of locality and causality are stated with respect to $M$. 

RQM focuses on relative variables describing systems ${\mathcal S}$ and ${\mathcal S}'$ with respect to each other, and as in standard quantum mechanics, the same variables also describe the mutual action 
between 
${\mathcal S}'$ and ${\mathcal S}$. 
In a spacetime setting it is more natural to focus on the {\em local action} of a subsystem on the rest of the system. 
We may consider two coupled fields, and describe their interaction, which would be described by relative variables locally depending on the two fields. 
Another situation to consider is a single nonlinear field, whose restriction to spacetime regions may be considered as subsystems locally interacting in the intersecting region. 

Many systems of physical interest are naturally described in frameworks with gauge freedom. For a system ${\mathcal S}$, let us also call ${\mathcal S}$ its space of histories according to the framework with gauge freedom. Then the space of physically distinguishable histories is not ${\mathcal S}$, but a quotient space that we will call $S/G$. It happens that when gauge systems, say ${\mathcal S}$ and ${\mathcal S}'$, couple to each other, the framework that appropriately describes their coupling is a new gauge system with total space of histories $S^T = S \bigotimes S'$%
\footnote{In the case of fields on vector bundles, they are sections of a new bundle with the same base and fibers that are the tensor product of the original fibers.} 
and space of physically distinguishable histories is $S^T/G^T$ which is not constructed by a product of $S/G$ and $S'/G'$. This has been known for a long time, but a very nice argument for the physical reason behind this fact is given in \cite{Rovelli:2013}. It turns out that nature is relational and systems that are relational have natural mathematical descriptions in terms of frameworks with gauge freedom, and they couple to each other following those rules. 

With the goal of being precise, we will follow the lead of Algebraic Quantum Field Theory (AQFT); see for example 
\cite{haag2012local}. AQFT captures a spacetime version of the Heisenberg picture for observables by associating to each spacetime region a local algebra of observables. These associations obey a set of axioms stating that local and causal relations between different spacetime regions are mapped to algebraic properties of the corresponding local algebras of observables; the resulting structure is called a causally local net of observables. Below we face the task of expressing these ideas in a relational language. 

We will present two independent points of view. Now we present a deconstruction point of view. Consider $S(U)$ the local algebra associated to spacetime region $U$ describing system ${\mathcal S}$. Its elements are the observables localizable at $U$. As we mentioned before, observables play the dual role of describing properties of a system and also describing actions on a system. In AQFT there is no subject; the system is acted upon, but the acting agent is never mentioned. The only thing that AQFT records is when-where ${\mathcal S}$ received the action. It may a useful abstraction to think that it does not matter who acted on the system as long as it has the same effect. (We will not take that point of view.) 
Additionally, the physical context should clarify the issue and tell us who acted on the system. We will appeal to contextuality. In a situation in which we have two coupled fields $(\phi, \psi)$, we may assume that $\phi$ was locally acted upon by $\psi$ and vice versa. This will be done within a context where the description takes place in ``the interaction picture'' following \cite{Fewster:2020}; we give a more detailed explanation in Section \ref{Measurement/InteractionInRQP}. In a situation where there is a single field $\phi$ and a spacetime region $V$ is used to define $\phi|_{\overline{V}}$ the system of interest being described with respect to the rest of the system $\phi|_{\overline{M \setminus V}}$, we may assume that $\phi|_{\overline{V}}$ is being locally acted upon by the rest of the system (in $\partial V$ their interaction region). 
Thus, when in AQFT they write $S(U)$ we may think that this is an abstraction for ${\mathcal S}\text-{\mathcal S}'(U)$ the algebra of relational ${\mathcal S}\text-{\mathcal S}'$ observables localizable at $U$. 

The other point of view considers two coupled spacetime systems ${\mathcal S}$, ${\mathcal S}'$, which together they make system ${\mathcal C}$. The algebraic description of the coupled system would determine a net of observables $\{ {\mathcal C}(U) \}_{U\subset M}$. Relational variables describing ${\mathcal S}$ with respect to ${\mathcal S}'$ are self-adjoint elements of ${\mathcal C}(U)$. 
One may consider that relational observables are only those which are invariant under isometries of the background spacetime $M$, and that non invariant information is regarded as gauge%
\footnote{Notice that the relative velocity between two particles is a vector and that if it is not zero it is not invariant under rotations. If you keep only the relative speed as a relational variable, a good reference system in dimension $d$ would have to consist of at least $d+1$ particles with independent velocities.}.
We may consider whether variables that are invariant under isometries are necessarily relational. The point of view now described starts with the definition that relational variables are those that are invariant under the appropriate group of gauge symmetries. This elegant and powerful point of view is used in the Perspective Neutral framework for Quantum Reference Frames \cite{delaHamette:2021oex}.

Wilsonian renormalization is not easy to find in the principles of AQFT. The physical interpretation of theories that fulfill AQFT's axioms is that they correspond to completely renormalized theories. That is, they could be free theories, exactly solvable models, or they may be constructed following the continuum limit of Wilsonian renormalization. In the latter case, they correspond to the organization of completely corrected effective theories at all possible measuring scales. 
More precisely, the limit involves an inverse limit (i.e. with arrows going in the opposite direction to the coarse graining maps) of effective theories and rescaling. 
Since all possible measuring scales are considered, the complete algebra associated to a nonempty spacetime region would not be finitely generated. 
For renormalizable theories, however, the Wilsonian continuum limit tell us that the algebra of observables can be approximated by finitely generated algebras after an appropriate rescaling and limiting procedure. 
This consequence of Wilsonian QFT is behind the results stating that the algebras of AQFT are von Neumann type $III_1$ factors; see  \cite{halvorson2006algebraicquantumfieldtheory} and references therein. 
\footnote{Additionally, there is an algebraic formulation of Wilsonian QFT implemented in a perturbative setting \cite{costello2011renormalization}. 
Also, Lattice Field Theory \cite{creutz1983quarks}, a nonperturbative implementation of Wilsonian QFT usually presented in a euclidian formulation that is amenable to Monte Carlo simulations, could naturally be presented in an algebraic language. }

Now we will argue that extending the ideas of RQM to a spacetime setting leads naturally to a Wilsonian point of view on QFT. Moreover, the operational philosophy of the relational point of view fits directly with the description of effective theories that have not yet been renormalized. The specific proposal that we justify below 
will take the form of a ``Finite Resolution Postulate'' in RQP.

In the original version of RQM \cite{Rovelli:1996} the Postulate of Limited Information stated 
``There is a maximum amount of relevant information that can be extracted from a system''. 

In the more recent presentation \cite{Rovelli:St2025}, that we follow in Section \ref{RQM}, the same proposal is presented with a change of emphasis. In that  reference,  the Postulate of Limited Information 
is described saying that 
``relevant information is finite for a system with compact phase space''. 

Since in field theory phase spaces are always infinite dimensional, 
it is clear that the idea behind the Postulate of Limited Information 
needs to be appropriately generalized before applying it to spacetime systems. 

The appropriate generalization is suggested by Wilsonian QFT. Let us start considering a laboratory situation where the system of interest ${\mathcal S}$ is a field. The experimentalist sets up a collection of measuring devices, filters and controllers. The measurements that take place during the experiment are the result of actions of a system ${\mathcal S}’$ on ${\mathcal S}$ controlled by the experimentalist. The controlled environment crafted by the experimentalist may be mathematically described as a set of observables ${\mathcal S}\text-{\mathcal S}'(U, \lambda)$, where $U$ is the spacetime region in which the experiment takes place and $\lambda$ is a label that lets us distinguish one laboratory set up from another one. 
Observables in ${\mathcal S}\text-{\mathcal S}'(U, \lambda)$ correspond to the selfadjoint elements of a star algebra $\widetilde{{\mathcal S}\text-{\mathcal S}'}(U, \lambda)$. The physical situation demands two important properties in an acceptable algebra $\widetilde{{\mathcal S}\text-{\mathcal S}'}(U, \lambda)$: (i) It needs to be finitely generated. (ii) Its elements need to reflect the fact that the experimentalist resources are finite; ${\mathcal S}’$ does not act on ${\mathcal S}$ with infinite energy or, equivalently, ${\mathcal S}’$ cannot act on ${\mathcal S}$ sharply measuring properties at points. 

We propose to condense both required properties into one saying that “the relational information contained in a quantum event is finite”. 
Let us explain. In the laboratory situation described above, quantum events result from the interaction between ${\mathcal S}'$ and ${\mathcal S}$ as described by some element of ${\mathcal S}\text-{\mathcal S}'(U, \lambda)$. 
The defining property of the event is that the “responsible variable” has definite evaluation. Thus, the event contains the following relational information: (I) the element of ${\mathcal S}\text-{\mathcal S}'(U, \lambda)$ that caused the event, and (II) the value that the variable takes at the event. 
Requiring that the relational information contained in a quantum event is finite we constrain the amount of information of types (I) and (II), which implies that the required properties hold. 

Let us clarify what we mean by ``relational information'' using an example. Consider that ${\mathcal S}$ is electromagnetic radiation and ${\mathcal S}'$ is a field describing a (monochromatic) photographic plate. 
A monochromatic photographic plate could be seen as a device measuring the position within a screen, but it would be more accurate to describe it as an array of detectors asking the electromagnetic radiation ``is your energy $h\nu$?''. 
When the energy has that value, the detector has a certain probability of detecting it producing a quantum event. 
After the plate detects a photon, the position in the screen is recorded in two ways: First of all the particular detector of the array can be identified. We consider this as {\em relational information} because it refers to the relation between ${\mathcal S}$ and ${\mathcal S}'$. 
{\em Independently}, the location of the detector can 
be given with arbitrary precision, but we do not consider that information to be relational. 
It refers to the relation between (the state of) ${\mathcal S}'$ and spacetime. The initial preparation of the state of ${\mathcal S}'$ is what keeps the information of exactly where the experimentalist placed the photographic plate (the array of detectors).%
\footnote{That important information determines the precise elements that comprise ${\mathcal S}\text-{\mathcal S}'(U, \lambda)$; if the photographic plate had been placed in a slightly different position the set of available relational variables ${\mathcal S}\text-{\mathcal S}'(U, \lambda)$ would be different. }
Since it is not information 
directly associated to the ${\mathcal S}\text-{\mathcal S}'$ interaction leading to the quantum event, we do not classify it as relational information contained in the event. 
We consider that the relational information contained in the event includes the element of the set of observables ${\mathcal S}\text-{\mathcal S}'(U, \lambda)$ (up to rescaling) that triggered the event and the evaluation obtained (which in this example was only a yes or no value). 

It seems that 
the laboratory situation described above is too restrictive as to extract lessons for QFT in general situations from it, 
but the continuum limit is the tool that lets us extrapolate. 
In the continuum limit of Wilsonian QFT as implemented in \cite{Manrique:2006} the observables of any possible laboratory setup are considered. 
Two basic ingredients are involved in the continuum limit. Regularization maps bringing any possible laboratory setup to each measuring scale. The set of subalgebras ${\mathcal S}\text-{\mathcal S}'(U, \lambda)$, which in \cite{Manrique:2006} are called measuring scales, is a partially ordered directed set where the notion of continuum limit makes sense. 
For the continuum limit to work, two limiting processes need to converge. 
First, an arbitrary observable has a collection of substitutes, one for each measuring scale. There is a theory at each scale, and if the predictions determined by those theories regarding the substitute observables converge, we have prediction for the original observable. 
Second, the theory corresponding to a given measuring scale is constructed as a limit of effective theories. Those effective theories use only finitely many degrees of freedom; in other words, a cutoff is placed making their space of histories finite dimensional. Each effective theory has adjustable parameters (``coupling constants”) which are fine tuned according to a renormalization prescription, which intends to “make them all model the same physics at large scales”. 
There is a notion of removing the cutoff, which at the same time allows for more and more degrees of freedom into the description and gets rid of the arbitrariness in the construction of effective theories. 
If the descriptions converge as the cutoff is removed, the limit defined a “completely corrected theory”, which is the one used in the first limiting process of the continuum limit. For details and the example of an interacting relativistic QFT constructed following this procedure see \cite{Manrique:2006}.

Does the statement “the relational information contained in a quantum event is finite” survive the continuum limit? The term “continuum limit” suggests that the infinite number of degrees of freedom are considered, and that this implies that a real event contains infinite information, but this intuition is not correct. The moral of the Wilsonian continuum limit is that any physical situation may be described by a finitely generated algebra at some degree of approximation. The same situation may be described at different measuring scales, and as the measuring scale is refined the descriptions may converge. Convergence is the non trivial test known as the renormalizability of the theory (in the Wilsonian sense). Convergence means that the predicted qualitative behavior of the system stabilizes and the numerical values of predictions converge. Thus, statements like “the relational information contained in a quantum event is finite” will survive the continuum limit, when the limit exists.

Then we propose that for spacetime systems the Postulate of Limited Information of RQM should be generalized as follows. 

\medskip

\noindent {\bf Finite Resolution Postulate:}\\ 
``The relational information contained in a quantum event is finite.'' 

\medskip

Now that we stated an element of RQP as a postulate, it seems appropriate to include the cornerstone of the whole approach also as a postulate (that in a logical presentation would be written before). 

\medskip

\noindent {\bf Sparse Event Ontology Postulate:}\\ 

``The set of quantum events 
associated with the interaction between two systems is a discrete set.''

\medskip

Notice that in laboratory situations the locus of interaction can be confined to a compact set $K$. Then, the Sparse Event Ontology Postulate implies that the set of events associated to the ${\mathcal S}\text-{\mathcal S}'$ interaction is a finite set. 
Then the Finite Resolution Postulate implies that the entire set of events that can take place in an experiment contain finite relational information.

\section{``Measurement''/interaction according to RQP}
\label{Measurement/InteractionInRQP}

A laboratory situation in which a measurement takes place can be described using a relational language. In QFT the literature dealing with the fundamental interpretational aspects of measurement is not broad. In the first part of this section we will describe the measurement framework presented by Fewster and Verch \cite{Fewster:2020} from the point of view of RQP (from the perspective of the experimentalist). 
We will close the section with remarks regarding a possible 
parallel study of the same situation describing the system of interest directly with respect to the probe system ignoring the experimentalist. 
We will focus on two issues. First we will discuss the nature of quantum events in a purely relational description. 
Then we will argue that the Finite Resolution Postulate 
favors a relational description with a Wilsonian point of view.

Consider the situation described in \cite{Fewster:2020} where there is ${\mathcal A}$ a system of interest, ${\mathcal B}$ a probe system and an experimentalist that after the ${\mathcal A}-{\mathcal B}$ interaction takes place measures an aspect of ${\mathcal B}$ (say the position of the pointer of the apparatus) in a quantum mechanical sense. The assumptions about the systems are that ${\mathcal A}$ and ${\mathcal B}$ are fields following the axioms of AQFT%
\footnote{Fewster and Verch use variation of the AQFT axioms; for details see \cite{Fewster:2020}.} 
which interact only within a compact region $K$. 
Another assumption is that the experimentalist has independently prepared systems ${\mathcal A}$ and ${\mathcal B}$ before their interaction in states $\omega, \sigma$ respectively. 

The point of view in \cite{Fewster:2020} regarding states coincides with RQP's as described in Section \ref{RQM}: The state is a bookkeeping device. 
The bookkeeping log is used to calculate probabilities of possible evaluations as conditional to be consistent with the log (and according to  the known states $\omega, \sigma$ in the past of $K$). 
Notice that a bookkeeping device when time is the only independent variable, as in QM, is simply an ordered list, and  for spacetime systems the bookkeeping needs to associate events to spacetime loci. 

Let us be more precise when we talk about the observables that describe the measurement. 
Systems ${\mathcal A}$ and ${\mathcal B}$ interact inside $K$ and behave as independent systems outside $K$. One option is to describe the interaction using relational variables of the coupled system ${\mathcal C}$. This is not only difficult, but it is not readily useful in the given laboratory situation. Instead, \cite{Fewster:2020} describes the interaction using variables in ${\mathcal U}$, a hypothetical uncoupled system that behaves exactly as the coupled system within spacetime regions that do not intersect $K$. 
This point of view it is adequate to describe ${\mathcal A}$ with respect to ${\mathcal B}$ (as if it were not coupled to ${\mathcal A}$). 
At the mathematical level, for regions $V$ that do not intersect $K$ there is a canonical isomorphism between the algebras ${\mathcal C}(V)$ and ${\mathcal U}(V)= {\mathcal A}(V) \bigotimes {\mathcal B}(V)$. 
The measurement of a property of ${\mathcal B}$ by the experimentalist is taken as associated to the variable $1 \bigotimes B \in {\mathcal U}(V)$ for some region $V$ in the future of $K$. 
Notice that the same variable has a corresponding element 
$O_B = \gamma^+ (1 \bigotimes B) \in {\mathcal C}(V)$ the coupled system, where $\gamma^+$ is the canonical isomorphism between the uncoupled and coupled systems for regions in the future of $K$. 
Then we see that 
$O_B$ is the Heisenberg evolution of a variable  describing the interaction of ${\mathcal A}$ and ${\mathcal B}$ at the interaction region $K$. 
The evaluation $\lambda$ that the experimentalist reads and writes in the log 
corresponds to observable $O_B$ of the coupled system that interacted at $K$, and provides information about system ${\mathcal A}$ described with respect to its interaction with ${\mathcal B}$ (from the perspective of the experimentalist). 

Fewster and Verch also find an ``induced variable'' $A \in {\mathcal A}(V')$ (for a region $V’$ in the past of $K$) that has the property of having the same expectation value (in state $\omega$) as $O_B = \gamma^+ (1 \bigotimes B)$ has in the state of ${\mathcal C}$ corresponding to the uncoupled state $\omega \bigotimes \sigma$ of ${\mathcal U}$. They do not claim that $O_B= \gamma^- (A \bigotimes 1)$. This helps them in the construction of a ``state update rule''.

The mathematical setting of Fewster and Verch gives us the information that the interaction took place within $K$. In \cite{Fewster:2020} they show that if a set of measurement variables of the same type are considered 
(i.e. observable $O’$ to take value $\lambda’$ due to interaction within a compact region $K’$, etc) 
the corresponding expectation values calculated as conditional to the other measurements in the log 
satisfy the properties that one could expect from the causal structure in $M$. For details see \cite{Fewster:2020}.

Let us now give a list of remarks reading the measurement framework according to RQP. 
We will focus on issues that would be critical in describing the same ``measurement''/interaction situation from a purely relational point of view. 
This exercise will help us refine this interpretation of QFT before we extend its principles to more general situations: \\

(i) {\em About the event triggered by $O_B$.} 
As mentioned above, in \cite{Fewster:2020} it is proven that the measurement described only affects other measurements of the same type if the other coupling region is causally in the future of $K$ 
(unless there is a correlation of the induced observables according to $\omega$). 
Thus, we do not know the precise location of the event triggered by $O_B$, but we know that it takes place inside $K$. 
We could try to find a probability distribution for its location inside $K$ following a spacetime analog of the procedure used by Rovelli to solve the problem of finding the moment in time when a event in RQM takes place \cite{Rovelli:1998}. 
The procedure of \cite{Rovelli:1998} can be followed to find the moment in time when an event takes place, but the issue of location in spacetime requires further localization: Recall the example of the photographic plate described in Section \ref{RQPisWilsonian}. 
There we argued that 
the relational information contained in the event could be used to exactly determine which one of the detectors that conform the photographic plate detected the photon. 
Independently, the precise location of the plate could be determined; this was not considered to be relational information contained in the event. 
Thus, the objective of calculating a probability distribution for the location of the event from relational information 
makes sense only within that context. 
In our example, we could use relational information to calculate the probability that a given detector is activated, but we could not find its precise spacetime location from that information. 
Additionally, if we know what detector was activated, and we also knew the exact position of the plate, 
would that provide a sharp location for the event? 
No. Even if the detector in the photographic plate is very small, it is not a mathematical point. 
Extrapolating from the example to the general situation requires an interpretation of the relational variables available at a given interaction scale. Their interpretation is to provide coarse grained information about system ${\mathcal S}$ with respect to system ${\mathcal S}'$. Even when coarse graining is modeled by decimation, the interpretation of the variables is to provide a representative sample. {\em Thus, with relational information at a given interaction scale the event can only be located in a coarse grained context: i.e. its location is fuzzy} (and determinable only at the probabilistic level as usual in quantum physics).

From the point of view where ``relational'' is taken to mean invariant with respect to the relevant automorphism group 
\cite{delaHamette:2021oex}, which in this situation is isometries of Minkowski space, the position of the plate is clearly not relational information.

(ii) {\em About the complete relational information content of the experiment and Wilsonian QFT.} 
The Sparse Event Ontology Postulate, The Finite Resolution Postulate and the compactness of $K$ imply that the set of events that take place due to the interaction at $K$ contains a finite amount of relational information. Thus, the interaction scale ${\mathcal A}\text- {\mathcal B}(K, \lambda)$ obtained as the largest common subalgebra of the set of algebras associated to regions containing $K$ is finitely generated and the set of evaluations of any of its variables is finite. 
Then it is natural to describe this situation using an effective theory using a type $I$ algebra. It may be the case that the effective theory can be completely corrected, and that a continuum limit  yields a concrete procedure to approach the type $III_1$ algebra, describing the continuum limit by means of type $I$ algebras. We only want to point out that following this procedure is natural in a purely relational framework since 
${\mathcal A}\text- {\mathcal B}(K, \lambda)$ happens to be finitely generated in all laboratory situations.

For the particular laboratory situation the experimentalist can tell us what set of observables she has access to; we know that it must be a subset of the set of selfadjoint elements of ${\mathcal C}(K)$ and that all of these variables are of the type $O_B$ for some $B \in {\mathcal B}(V)$ for a region $V$ in the future of $K$. 

In the description of the interaction with respect to the probe system (where the experimentalist is not considered), the elements of ${\mathcal C}(K)$ that should be called relational variables are not of the same type.
Relational variables correspond to mutual actions between ${\mathcal A}$ and ${\mathcal B}$. If we want to describe them from the point of view of the hypothetical uncoupled system (where we still assume that interaction is confined to a compact set $K$), the variables should be a sum of terms of the type $\gamma(A_i \bigotimes B)_i$, where the pairs $(A_i, B_i)$ describe reciprocal actions. 
From the point of view where relational is taken to mean invariant with respect isometries of Minkowski space, 
the allowed variables should be invariant under Lorentz transformations acting on ${\mathcal U}(K)$.

\section{Summary and outlook}

As we emphasized in the introductory section, for interacting QFTs Wilsonian renormalization is necessary. This is very different from the ``accidental situation of QM'' which considers a 1 dimensional spacetime. 
Spacetimes of dimension 1 have Cauchy surfaces of dimension 0, which means that they have finite dimensional spaces of initial conditions in contrast to what happens in higher dimensions. 

In this paper, motivated by the Wilsonian view of QFT, we propose to generalize the 
Postulate of Limited Information of the original formulation of RQM to apply to spacetime systems in the form of the Finite Resolution Postulate. 
The form of the postulate fits so naturally in the relational view of RQM that it is tempting to say that the relational point of view of quantum physics leads to the cornerstone of Wilsonian QFT. Our presentation in Sections \ref{RQPisWilsonian} and \ref{Measurement/InteractionInRQP} provides supporting elements. 

We argued in Section \ref{Measurement/InteractionInRQP} that an event that took place at a given interaction scale cannot be sharply located by means of relational information. 
Similarly, the prediction of a probability distribution for the location of future quantum events can only be done with the resolution available at the relevant interaction scale. 
This is an important difference with what happens in QM, where quantum events are located in time sharply.

An important element of the measuring scheme of Fewster and Verch \cite{Fewster:2020} is to consider a situation in which the interaction among systems ${\mathcal S}$ and ${\mathcal S}'$ is confined to a compact set $K$. 
That situation made possible a description of the coupled system in terms of a hypothetical uncoupled system and a scattering map. 
Clearly, the general situation for interacting fields is not of this type. 
We have plans to address that problem in the near future. As a spoiler, we are modeling the interaction region as the gluing of cells. 
The compact dimension full cells host a coupled system and a hypotetical uncoupled system. Thus, the dynamics of the coupled system can be casted in terms of a ``scattering map'' acting on the restriction of the uncoupled system  to the boundary. 
Then the gluing of neighboring higher dimensional cells 
to model larger domains of interaction 
is done, at the level of the uncoupled systems, by the imposition of appropriate gluing equations.

In the introduction we mentioned that gauge systems are among the most important systems in physics. 
We also mentioned that the coupling of this type of systems is described at the non gauge invariant level. 
The reason for their importance in physics and for the way in which they couple to each other 
seems to be tied to physics being essentially relational \cite{rovelli2014gauge}. 
Then work on the relational interpretation of quantum physics should address gauge systems. We plan to address this issue together with the problem of extending to non confined interaction domains. 

The Perspective Neutral Framework of \cite{delaHamette:2021oex} and the Positive Formalism of the General Boundary Formulation 
\cite{Oeckl:2005bv, Oeckl:2016tlj, Oeckl:2025gcb} treat quantum fields with the goal of being relational in some way. It would be interesting to 
take into account the Finite Resolution Postulate and use it as the initial ingredient to incorporate Wilsonian renormalization into those frameworks. 

The framework known as quantum reference frames shares motivations with our work. 
It would be interesting to see if our manifestly spatiotemporal point of view leads to important differences in the issues addressed in \cite{Castro-Ruiz:2019nnl, Guerin:2018fja}.

\medskip

Supported by grant PAPITT-UNAM IN114723 

\medskip
 
\bibliographystyle{unsrt}

\bibliography{bibliography}

\end{document}